**Hans C. van Houwelingen[1], Ronald Brand**

Department of Medical Statistics, Leiden University Medical Center,

Leiden, The Netherlands

**Thomas A. Louis**

Division of Biostatistics, School of Public Health

University of Minnesota, Minneapolis, USA


**Abstract**


The paper discusses empirical Bayes methodology for repeated quality comparisons of health care institutions using data from the Dutch VOKS study that annually monitors the relative performance and quality of nearly all (about 140) Dutch gynecological centres with respect to different aspects (interventions and outcomes) of the childbirths taking place in these centres.

This paper can be seen as an extension of the pioneering work of Thomas, Longford and Rolph [20] and Goldstein and Spiegelhalter [7]. First of all, this paper introduces a new simple crude estimate of the centre effect in a logistic regression setting. Next, a simple estimate of the expected percentile of a centre given all data and a measure of "rankability" of the centres based on the expected percentiles are presented. Finally, the temporal dimension is explored and methods are discussed to predict next year's performance.



[1]Corresponding author. Address: P.O. Box 9604, 2300 RC Leiden, The Netherlands; phone: +31-71-5276826; fax: +31-71-5276799; email: jcvanhouwelingen@lumc.nl




## 1. Introduction

Quality comparison of health care and educational institutions has drawn much attention over the last years. Statisticians have picked up the issue in an early stage and the awareness has grown that such comparisons ask for proper statistical methodology. See the commentary by Gore [8], the very nice paper by Spiegelhalter [19] and the editorials of McKee [13] and McKee and Sheldon [14] in the British Medical Journal.

Statistically speaking, there are two stages in dealing with the comparison problem. First, the difference in patient mix between institutions should be taken into consideration and corrections should be made for those differences by proper regression models. That stage is relatively easy to carry out, although it can be hard to deal with the pitfalls of observational modelling such as selection bias. It yields some crude estimates of performance for all institutions with some measure of imprecision. The second, much harder, stage is to draw proper conclusions from these crude estimates. In the clinical literature, there is an irrepressible tendency to rank institutions and to produce so-called league tables as if the institutions were taking part in some sort of a competition. A nice example of such tables is given in the recent paper by Parry et al. [16].

The statistical community agrees that empirical Bayes models can be very useful here to give a more realistic description of the differences between the institutions. ( However, it does not agree about the terminology; terms as hierarchical model and multi-level model are used to describe the same model). Thomas, Longford et al. [20] introduced the empirical Bayes models in this setting. The paper by Goldstein and Spiegelhalter [7] sketching the merits of the hierarchical model has been very influential and has found its way into practical applications as Marshall and Spiegelhalter [12] and Leyland and Body [10]. Refinements and



extensions are discussed in Normand, Glickman et al. [15], Deely and Smith [4] and Daniels and Gatsonis [3]. DeLong et al. [5] compare fixed effects models with random effects models. Cox [2] mentions the problem in his broad overview of the current position of statistics.

This paper arises from a Dutch project on the quality comparison of gynecological units with respect to different aspects of childbirth, in which the Leiden Department of Medical Statistics has been involved for a long time. In this project all participating centres receive a yearly report on their performances. Hence, we have a longitudinal series of yearly comparisons as in Parry et al. [16]. As argued in Van Houwelingen [21], the interesting issue here is how well the relative position of an institution for next year can be predicted. If so, that asks for measures by the authorities to correct "bad predictions". If not, the whole exercise does not make very much sense.

The main novel issue of this paper is how to deal with longitudinal data in the context of "league tables". Along the line, a new crude estimate of the centre effect will be presented in the line of Yusuf-Peto approach in meta analysis [22]. Furthermore, the ranking problem will be reviewed and a modification of the expected percentile, going back to Laird and Louis [9], see also Shen and Louis [18], will be advocated as a very suitable summary measure.

Throughout the paper we will use the "classical" plug-in empirical Bayes methodology. The advantage is that it can easily be implemented in software packages as SAS PROC MIXED, but we do realize that it ignores the uncertainty in the parameters of the mixing distribution and, therefore, may lead to overrating the differences between the institutions.

The paper is organized as follows: Section 2 gives more information about the Dutch VOKS project, such as the number of institutions and patients, the relevant outcomes and the explanatory variables that describe the patient mix. Section 3 discusses crude estimates of the



centre effects from a likelihood perspective. Section 4 reviews the classical empirical Bayes model and the ranking problem. In section 5 some results will be given from the VOKS study illustrating the issues of sections 3 and 4. The extension of the model to longitudinal data is given in section 6 and results for the VOKS data in section 7. Finally, a concluding discussion is given in section 8.

## 2. Description of the problem and the VOKS data

The VOKS study (Verloskundige Onderlinge Kwaliteits Spiegeling) annually monitors the relative performance and quality of nearly all (about 140) Dutch gynecological centres with respect to different aspects (interventions and outcomes) of the childbirths taking place in these centres. The project started in 1988. The data from the first years are not very reliable. After 1995 an intervention study was started that randomized the centres into a group that received yearly reports on their own performance and a group that is not informed. Since the intervention study might influence the results, we only consider the data for the five year period 1991-95.

The infants are grouped into 4 subgroups: very preterm (25-32 weeks), preterm (32-37 weeks), at term (37-42 weeks) and postterm (42+ weeks). The annual sizes of these groups are of the order 1,500, 10,000, 80,000 and 6,000 respectively. This does not cover all childbirths in The Netherlands, because a substantial part off all deliveries take place at the mother's home under guidance of a midwife or the family doctor.

In the study several measures of intervention and patient's outcome are considered. In this paper we concentrate on two dichotomous measures: the intervention Caesarian Section (CS) and the outcome Mortality within seven days (M). An indication of the crude CS and M rates in the four subgroups is given in Table 1.



The original purpose of the project was to report to all participating centres their relative performance after correction for explanatory variables at the level of the infant/mother. No explanatory variables at the centre level (such as centre size) were taken into account.

Logistic regression models without centre effects were used to compute the expected "risk" for each centre. The explanatory variables describing the "patient mix" are

1.     gestational age
2.     birth weight
3.     lethal congenital malformation
4.     multiple pregnancy
5.     fetal position
6.     tension of the mother
7.     planned delivery at the institution or referred by a midwife or family doctor in a late stage
8.     duration of primary care
9.     intra-uterine transport
10.    ethical minority
11.    estimated mortality risk

In principle, all covariates are used in the modelling of CS. In the modelling of M only covariates 1-9 are used. Different functional models (inclusion and coding/transformation of covariates, inclusion of interactions) are used in the four subgroups. The same functional form of the model was used each year, but the coefficients were re-estimated each year. The difference between observed risk and expected risk as derived from the logistic regression model was taken as performance measure to assess the quality of an institution. The institutions were ranked on this criterion and received information about their own ranking on the different measures.

In the light of [7] and our own research the practice of ranking based on crude estimates has been abandoned. Empirical Bayes methodology has been introduced, using the original logistic regressions as starting point.



### 3. Obtaining crude estimates of the centre effects

We consider a dichotomous outcome $Y_{ijk}$ that has been measured in centre $i$, in year $j$ on patient $k$ $(i=1,...,I, j=1,...,J, k=1,...,n_{ij})$. (We use the term patient throughout in a broad sense. In the example patient stands for the infant/mother combination) We restrict attention to dichotomous outcomes and logistic regression, but the whole set up easily carries over to other generalized linear models and survival analysis.

The usual logistic regression model is

$$\text{logit}(P(Y_{ijk}=1|X_{ijk},\theta_{ij})) = X_{ijk}^{'}\beta + \theta_{ij} \qquad (3.1)$$

with

$X_{ijk}$      the vector of covariates describing the patient's characteristics and including the constant term,

$\beta$      the vector of regression coefficients describing the effect of the covariates and the intercept,

$\theta_{ij}$      the "effect" of the $i$-th centre in the $j$-th year, that is ln(odds ratio) with respect to some overall mean.

If there are many centres involved and some centres have small sample sizes, fitting such models may lead to exploding centre effects. This could either be remedied by using a mixed model with random $\theta_{ij}$'s model right from the start or by following a two step procedure as used in Thomas et al. [20].. Mixed models could be fitted by software packages as SAS GLIMMIX or ML-Win, but it is very time-consuming to find the best regression model in (3.1) in a mixed model context with many centres and many observations. We will follow the



two step strategy, partly because it is in line with the original analysis strategy of the VOKS project and also because it is easy to communicate. In the first stage, we estimate the regression coefficients $\beta$ ignoring the centre effects $\theta_{ij}$. In the sequel we concentrate completely on the $\theta_{ij}$'s and act as if the regression coefficients $\beta$ are known.

Define

i)   $O_{ij} = \sum_k Y_{ijk}$

ii)   $p_{ijk} = P(Y_{ijk} = 1) = \exp(X'_{ijk}\beta)/(1 + \exp(X'_{ijk}\beta))$   (3.2)

iii)   $E_{ij} = \sum_k p_{ijk} = \sum_k \exp(X'_{ijk}\beta)/(1 + \exp(X'_{ijk}\beta))$

iv)   $var_{ij} = \sum_k p_{ijk}(1 - p_{ijk})$

Obviously, a measure of performance of centre $i$ in year $j$ will somehow depend on the difference $O_{ij} - E_{ij}$. A popular measure, used and defended in Thomas et al. [20], used in Parry et al. [16], DeLong et al. [5] and, originally, in the VOKS study is

$$W_{ij} = (O_{ij} - E_{ij})/n_{ij}$$   (3.3)

with standard error

$$(1/n_{ij})\sqrt{var_{ij}}.$$   (3.4)

However, this measure is not quite in line with the logistic regression model. It is easy to show that the Taylor expansion of the log-likelihood $l_{ij}$ for $\theta_{ij}$ around $\theta_{ij} = 0$ is given by

$$l_{ij}(\theta) = l_{ij}(0) + (O_{ij} - E_{ij})\theta - \tfrac{1}{2}var_{ij}\theta^2 + \ldots$$   (3.5)

That suggests to use

$$\hat{\theta}_{ij} = (O_{ij} - E_{ij})/var_{ij}$$   (3.6)

as a crude measure of (the lack of) performance and to act as if



$$\hat{\theta}_{ij} \sim N(\theta_{ij}, 1/var_{ij}) \hspace{3cm} (3.7)$$

A similar approximation was introduced in the field of meta-analysis by Yusuf and Peto [22]. It is an application of the notion of the (efficient) score statistic. See Cox and Hinkley [1, chapter 9] for more theoretical background. The equations (3.6) and (3.7) are used in a slightly different context by Louis and Bailey [11].

Notice that the approximate normality in (3.7) not a statement about the sampling distribution of $\hat{\theta}_{ij}$, but one about the validity of using the Taylor expansion of the log-likelihood as function of $\theta_{ij}$ around 0. For the time being we will assume that the true centre effects are relatively small. We will check the validity of this assumption for our data later on.

In practice there is little difference between an analysis based on our $\hat{\theta}_{ij}$ and one based on $W_{ij}$. Roughly speaking $\hat{\theta}_{ij} \approx W_{ij}/(\bar{p}(1-\bar{p}))$ with $\bar{p}$ =average p. The whole analysis as presented in the sequel could also be based on $W_{ij}$. Using W might be more appealing to the intuition of the practician while $\hat{\theta}_{ij}$ appeals more to the theoretical mind. We prefer using $\hat{\theta}_{ij}$ because its comes much closer to doing a full scale Maximum Likelihood analysis for model (3.1). We also like the idea that the approximate normal distribution is not based on the Central Limit theorem but on a Taylor expansion of the likelihood. From now on, we act as if the regression coefficients contained in the vector $\beta$ are known, use (3.6) as our definition of performance measure and act as if (3.7) is true.

## 4. Univariate empirical Bayes model

For the time being we restrict attention to one fixed year and suppress the year index j. We want to make inference about the individual effects $\theta_i$ for each centre knowing that we have



estimates $\hat{\theta}_i$ that are approximately $N(\theta_i, s_i^2)$ (In the likelihood sense as discussed in section 3).  This information can be graphically displayed as a sequence of 95%- confidence intervals

$$95\%\text{-CI} = (\hat{\theta}_i \pm 1.96 s_i) \tag{4.1}$$

Using the crude estimates may lead to misleading conclusions, especially if the standard errors $s_i$ differ from centre to centre. Small centres tend to have large $s_i$ and extreme estimates $\hat{\theta}_i$.

At this stage we do not consider explanatory variables at the centre level.  Technically speaking that would not be very hard, but we stick to the simple situation that centre effects are only corrected for patients mix. We will come back to this point in the discussion.

As pointed out by Robinson [17], the empirical Bayes approach  is very well suited for handling this kind of data and it gives a much more realistic idea of  the unknown effects in many applications. The practical results of this approach might even convince those who have strong philosophical objections against declaring the unknown centre effects to be random. The empirical Bayes approach takes the centre effects $\theta_i$ to be random with some distribution G. The simplest model is $G = N(\mu, \tau^2)$. One might even postulate here that $\mu = 0$ because the intercept term is already contained in the regression part of the model,  but the non-linearity of the logistic model may ask for a $\mu \neq 0$. Simple estimates of $\mu$ and $\tau^2$ can be found in DerSimonian and Laird [6], who consider the same model in the context of meta-analysis. However, it is not hard to obtain the Maximum Likelihood Estimates for  $\mu$ and $\tau^2$ from the marginal $N(\mu, s_i^2 + \tau^2)$-distribution (using SAS PROC MIXED or similar software) or via a home-made EM-algorithm.  (The MLE might not be the best estimator for $\tau^2$. Alternatives are REML or some Bayesian estimator. The actual choice of the estimator for $\tau^2$



is not very crucial for the reasoning in the rest of the paper)

The interesting part for further inference about centre effects is the posterior distribution of $\theta_i$. It is given by

$$\theta_i|\hat{\theta}_i \sim N(EBE_i \ , \ pv_i) \tag{4.2}$$

with

$$EBE_i = \mu + \tau^2/(\tau^2 + s_i^2) * (\hat{\theta}_i - \mu) \tag{4.3}$$

the posterior mean.  and

$$pv_i = \tau^2 s_i^2/(\tau^2 + s_i^2) \tag{4.4}$$

 the posterior variance.

The posterior mean is known as the empirical Bayes estimate (EBE) of the centre effect.

The data can be graphically presented as a sequence of 95%-posterior probability intervals

$$95\%\text{-PPI} = (EBE_i \pm 1.96\sqrt{pv_i}) \tag{4.5}$$

It is interesting to reflect for a while about the differences between the fixed effect model with its confidence intervals (4.1) and the random effect model with it posterior probability intervals (4.5) . If all centres have the same standard error $s_i$, the two representations differ only in scale. The *EBE*'s are closer to the common mean $\mu$ , but (4.1) and (4.5) reflect the same uncertainty about the ranking of the centres.  If the centres have different standard errors $s_i$, the story is more complicated. If the standard error $s_i$ is small, the *EBE* is close to the MLE and the posterior variance close to $s_i^2$. If  $s_i$ is large,  the *EBE* is close to $\mu$ and the posterior variance close to $\tau^2$.



On the average, the *EBE*'s give better estimates of the true centre effects than the crude estimates. The posterior distribution summarizes our (lack of) knowledge about the centre effects. (Notice that the histogram of the *EBE*'s does not estimate the mixing distribution *G*. See Shen and Louis [8] for a discussion about how to combine estimating *G* and ranking the centres in a single procedure)

Ranking of centres according to quality is very popular. The simple approach, still used in Parry et al.[16] can be misleading because the result is very much affected by random variation and there is no room for the possibility that it is all pure random noise and ranking does not make sense at all. Ranking of the *EBE*'s is not much better because it also ignores the uncertainty reflected by the posterior variances $pv_i$.

Most authors agree that whatever way we want to rank the centres, it should be based on the posterior distributions (4.2). Goldstein and Spiegelhalter [7] simulate from the posterior distribution and compute the median rank and its Bayesian confidence interval by means of Markov Chain Monte Carlo (MCMC). Deely and Smith [4] advocate the use of a single number, such as $P(\theta_i < \mu | data)$ or $P(\theta_i = \min_j(\theta_j) | data)$.

We think that the expected rank, as proposed by Laird and Louis [9] is a very suitable summary measure. It is defined as

$$ER_i = 1 + \sum_{j \neq i} P(\theta_j < \theta_i | data) = 1 + \sum_{j \neq i} \Phi((EBE_i - EBE_j)/\sqrt{pv_i + pv_j}) \qquad (4.6)$$

It is very close to the median rank of Goldstein and Spiegelhalter [7]. There are two extreme possibilities. If the standardized differences $(EBE_i - EBE_j)/\sqrt{pv_i + pv_j}$ are very close to zero, that is if the posterior probability intervals show considerable overlap, the expected ranks *ER*



are all close to the mid-rank *(n+1)/2* . If the standardized differences are very large, that is if there is hardly any overlap between the *PPI*'s , the expected ranks get very close to the ranks based on the posterior means *EBE*.

It is tempting to rank the centres on the basis of the expected ranks $ER_i$, as suggested in Laird and Louis [9] and implicitly done in the graphics of Goldstein and Spiegelhalter [7]. However, such a ranking hides the underlying uncertainties and , therefore, we prefer to use the expected rank as such and not to transform them into ranks 1,..n,.

Ranks are a bit inconvenient in practice, because , by definition, they depend heavily on the number of centres. Therefore, we propose to compute percentiles instead defined by

$$PCER_i = 100 * (ER_i - 0.5)/n \qquad (4.7)$$

The symbol *PCER* stands for "PerCentiles based on Expected Ranks". This percentile can be interpreted as an estimate of the probability that the effect $\theta_i$ is smaller than the effect $\theta_j$ of an randomly selected centre. Here, selection is from all n centres included an independent copy of centre i itself.   It is easy to extend this notion to the whole population of all conceivable centres. That means that we want an estimate of the  percentile $100 * G(\theta_i)$ . A straightforward estimate is

$$EPC_i = 100 * E[G(\theta_i | data)] = 100 * \Phi((EBE_i - \mu)/\sqrt{\tau^2 + pv_i}) \qquad (4.8)$$

Here, the symbol *EPC* stands for "Expected Percentile".

In the practical examples discussed below, there turned out to be hardly any difference between *PCER* and *EPC*.  The latter has the advantage of a very simple computation. Therefore, we will base our presentation on *EPC*, but emphasize that there is very little conceptual difference between this new measure and the expected rank (*ER*) of Laird and



Louis [9] and the median rank of Goldstein and Spiegelhalter [7].

The variance of *EPC* can be compared with the maximal variance of uniform percentiles ($100^2/12$). The ratio can serve as a measure RA of "rankability". Formally, we define

$$RA = 12 * var(EPC)/100^2 \qquad (4.9)$$

This measure describes the true variation of the percentiles on a 0-1 scale. If it is very small, the differences between the centres are completed obscured by the measurement error. We could define a similar and even simpler measure based on the variance $\tau^2$ of the random effect and the total variation of the crude estimate. Here, we have to take account of the variability in the error variance $s_i^2$. Since the distribution of the error variance is rather skewed , we take the median as typical value. That leads to our measure of proportion true variation between centres

$$\rho = \tau^2/(\tau^2 + median(s_i^2)) \qquad (4.10)$$

It measures what part of the variation between the crude centre effects is due to true differences (as opposed to measurement error). It is the theoretical correlation between crude centre effects in different years if the true centre effect is constant over time. Therefore, we denote it by $\rho$.

## 5. Annual results for the VOKS data.

As mentioned before we will consider the data from the years 1991-95 on the endpoint Caesarian Section and Mortality in the subgroups Very Preterm, Preterm, At Term and Postterm. For each subgroup×year combination we fit the $N(\mu,\tau^2)$ mixture model to the crude



estimates of the centre effects defined by (3.6) using the likelihood of (3.7). Correlation between centre effects over the groups or over the years is ignored for the time being. The estimated parameters of the empirical Bayes model are very stable over the years. Table 1 gives some typical numbers for each of the subgroups.

Table 1. about here.

The fitted values of $\mu$ (not shown because they are not very relevant) and $\tau^2$ indicate that the true centre effects are indeed close to zero and the approximation (3.7) seems reasonable here.

The variance $\tau^2$ of the random centre effects is largest in the first subgroup of Very Preterm infants for both outcomes. However, due to the small number of patients the variability is very large and the proportion true variation $\rho$ quite small. In the other subgroups the mortality rates are very low, the variance component $\tau^2$ and the proportion true variation $\rho$ are very small, so it seems that any attempt to discriminate between the centres on that outcome in these subgroups is hopeless. For the outcome Caesarian Section, the variance component $\tau^2$ is quite small, but the proportion true variation $\rho$ is much higher due to the high Caesarian Section rates in all these subgroups. The proportion true variation is maximal in the At Term subgroup due to the large number of patients.

To get more insight we consider two situations in more detail. First of all, the outcome Caesarian Section in the At Term group in 1995. That should be about the ideal case for comparison of the centres. Second, we consider the outcome Mortality in the subgroup Very Preterm in 1993, because in that year the variance component $\tau^2$ was the largest of all years. That should be a borderline case for any sensible comparison of the centres.



Caesarian Section in the At Term group in 1995.

There are 112 centres The mean number of patients per centre is 695 and the overall CS rate is 16%. The confidence intervals of equation (4.1) are plotted in figure 1.

Figure 1 about here..

The estimated parameters of the random effect distribution are $\mu = 0.038$ and $\tau^2 = 0.124 (\tau = 0.352)$. The proportion true variation $\rho = 0.91$. The posterior probability intervals of (4.5) are shown in Figure 2 and the scatter plot of the expected percentile *EPC* versus the crude percentile in Figure 3.

Figure 2 and Figure 3 about here.

The measure of rankability *RA*=0.90, very close to the measure of proportion true variation $\rho = 0.91$. The conclusion in this subgroup is that, thanks to the large sample sizes per centre and the high CS rate, it very well possible to discriminate between the centres although the actual differences are relatively small.

Mortality in the group of very preterm infants in 1993

There are 112 centres. The mean number of patients per centre is only 14.3, The overall mortality rate is 26%. Figure 4 shows the very noisy 95%-confidence intervals

Figure 4 about here



The random centre effects have mean $\mu=0.23$ and variance $\tau^2=0.336$ ($\tau=0.579$). The

proportion true variation $\rho=0.19$. Figure 5 shows the 95%-posterior probability intervals. It is

clear that the empirical Bayes methodology scales down the estimated centre effects to much

more realistic proportions but it does not help very much in producing any reliable ranking of

the centres. That is quantified by the low rankability index RA=0.23. Figure 6 shows the

crude percentile, the percentile based on the *EBE*'s and the *EPC*. First, it shows that the order

is very much changed going from crude estimates to *EBE*'s. Second, it shows that the vast

majority of the centres have an expected percentile between 30% and 70%. There are a few

(larger) centres that have an outspoken low risk.

Figures 5 and 6

The conclusion for this subgroup must be that, due to the small sample sizes per centre, it is

very hard to make any inference on the relative quality of an individual centre, although there

is a substantial (latent) variation between the centres reflected by the relatively large variance

component $\tau^2$.

## 6. Extension of the model to longitudinal data

The whole exercise of comparing centres in a particular year only makes sense if the results

of that year carry over to the next year. If we have J annually repeated measures as in the

VOKS data set, we can both check the correlation between the years and produce some

estimate for the performance in year J+1 by using an appropriate mixed model for the

repeated measures. As for the univariate case, the model could best be described as a two

stage model



$$\hat{\Theta}_i | \Theta_i \sim MVN(\Theta_i, S_i) \qquad\qquad (6.1)$$

$$\Theta_i \sim MVN(M, T)$$

The vector $\hat{\Theta}_i$ contains the crude estimates for centre $i$ in each year . Conditional on the true

centre effect $\Theta_i$, these estimates are multivariate normal with mean $\Theta_i$ and diagonal

covariance matrix $S_i$ containing the conditional variances $s^2$ of section 4. The true centre

effects have a multivariate normal distribution with mean M containing the µ 's of section 4

and covariance matrix T containing the annual variance components $\tau^2$ on the diagonal and

allowing correlation between the years. Fitting such a model to the data is not extremely hard.

The EM algorithm is very suited for fitting the two-stage model, but it is a bit slow. Standard

software as SAS Proc Mixed could be used as well. We will give some more detail in the

next section. Missing data could occur if a centre joins the exercise in a later stage or drops

out, but such missing data does not constitute a serious problem as long as it is missing at

random (MAR). Fitting some model to the observed data gives an understanding of the

correlations between the years and the (im)possibility to say anything about next year. If we

really want to make any statement about the next year we need to extrapolate both M and T.

The extrapolation of M is not crucial because the actual mean µ is not relevant in measures as

*EPC* or *RA*. However, it is crucial to extrapolate T. Models for the covariance matrix that

allow easy extrapolation are the compound symmetry model (constant variance, equal

correlations), the random regression coefficient model $\Theta_j = ZB_i$ with $Z_1 = constant$ and

$Z_2 = {}^{//}time{}^{//}$ leading to $T = Z * cov(B) * Z^{/}$ , the pure stationary autoregressive model $T = \tau^2 R$



with $R_{ij} = \rho^{-|i-j|}$, or a combination of a random intercept and an autoregressive term. We will discuss some models in more detail in the next section. Once the model is fitted on the data from the years $1,..,J$ and $M$ and $T$ are properly extrapolated, it is not hard to compute the distribution of $\theta_{i,J+1}$ given all data of centre $i$. It is normal with

$$E[\theta_{i(J+1)}|\hat{\theta}_{i1},...,\hat{\theta}_{iJ}] =$$

$$\mu_{J+1} + \begin{pmatrix} T_{1,J+1} \\ ... \\ T_{J,J+1} \end{pmatrix}' \begin{pmatrix} T_{11}+s_1^2 & ... & T_{1J} \\ ... & ... & ... \\ T_{J1} & ... & T_{JJ}+s_J^2 \end{pmatrix}^{(-1)} \begin{pmatrix} \hat{\theta}_{i1}-\mu_1 \\ ... \\ \hat{\theta}_{iJ}-\mu_J \end{pmatrix}$$

$$(6.2)$$

$$var[\theta_{i(J+1)}|\hat{\theta}_{i1},...,\hat{\theta}_{iJ}] =$$

$$\tau_{J+1}^2 - \begin{pmatrix} T_{1,J+1} \\ ... \\ T_{J,J+1} \end{pmatrix}' \begin{pmatrix} T_{11}+s_1^2 & ... & T_{1J} \\ ... & ... & ... \\ T_{J1} & ... & T_{JJ}+s_J^2 \end{pmatrix}^{(-1)} \begin{pmatrix} T_{1,J+1} \\ ... \\ T_{J,J+1} \end{pmatrix}$$

Similar expressions hold if the past information is not complete.

Having obtained these conditional predictive distributions, the predictive rankability of the centres for year J+1 can be further analysed in the way of section 5. We will demonstrate that in the next section.

## 7. Longitudinal results for the VOKS data



Continuing the example we considered before, we first fit a saturated (unstructured) mixed model for the outcome Caesarian Section in the At Term group and the outcome Mortality in the Very Preterm group. Unstructured models with missing data are very easy to fit by the EM algorithm. We used a hand made EM algorithm written in Gauss. The results are given in Tables 2 and 3.

Insert Tables 2 and 3 about here

In the Caesarian Section data we see very high correlations between the true centre effects over the years. Those correlations could already be observed from the crude estimates (not shown). So we can conclude that the centre effects are quite stable over the years and we trust that the centre effects for the next year are quite well predictable. The Mortality data in the Very Preterm group are in some sense much more interesting because there is so much noise and the centre effects are very hard to estimate. The correlations between the true centre effects in Table 3 are surprisingly high and consistent. In the crude data these correlations are completely hidden by the large amount of noise. We will elaborate these data in more detail to see how predictable the centre effects are. We fit two different models and compute for each centre the predictive distribution in both models using all available data from the past.

Model I is a stationary autoregressive model with covariance matrix $T = \tau^2 R$ with $R_{ij} = \rho^{-|i-j|}$

and unstructured mean. The correlation parameter $\rho$ is a bit hard to estimate. Given that parameter, estimation of the mean and variance parameters can easily be done by EM. So we estimated $\rho$ by maximizing its profile likelihood derived from the EM algorithm for the other parameters. We did not model the mean value, but extrapolated it manually to the next year. Model II is the random coefficients model



$$\theta_{ij} = A_i + B_i * (year_j - 90)$$

with

$$\begin{pmatrix} A_i \\ B_i \end{pmatrix} \sim N \left( \begin{pmatrix} \alpha \\ \beta \end{pmatrix}, \begin{pmatrix} \tau_A^2 & \rho_{AB}\tau_A\tau_B \\ \rho_{AB}\tau_A\tau_B & \tau_B^2 \end{pmatrix} \right)$$

Again, this model is well suited for fitting by the EM-algorithm.

The results of the model fit are given in Table 4.

Insert Table 4

The goodness-of-fit of both models is very similar, the random regression coefficient model fits slightly better, presumable because it adapts to the slowly increasing yearly variances in Table 3. On the other hand, extrapolations of model I are more robust than those of model II. Next, we compute the predictive distribution for 1996 for each centre using both models. As can be seen from formula (6.2) predictive distribution does not only depend on the model we use for the true centre effects and the past crude estimates for each centre, but also on the precision of the crude estimates that directly depend on the size of the centre. From the predictive distributions the expected percentile (4.8) is computed for each centre and the overall predicted rankability RA (4.9). Notice that the concept of the proportion true variation does not carry over easily to the prediction setting while the concept rankability does.

The two prediction models are not completely identical. There is some shift in scale and location. The results for the better fitting model II are shown in Figure 7. The picture for model I looks very much the same. The Expected Percentiles according to the two models are very similar as can be seen in Figure 8.



Figures 7 and 8 about here.

The rankability in Figure 7 is $RA = 0.38$. Comparing this with section 4 where we found $RA = 0.23$, we can say that it is easier to predict the rankings for the year 1996 using the data from the period 1991-95 than to determine the rankings in the year 1993 using the data from that year alone. This shows what high correlations over time can do to improve predictions and that surveillance of the centre effects make sense even in the case of small centres with few patients, provided the true effects show a consistent pattern over time. For the data considered the consistency over time could already be concluded from the correlations in Table 3.

## 8. Discussion

The results of this paper show once again that the empirical Bayes framework is very well suited for the analysis of quality comparison data. We think that our Expected Percentile measure (4.8.) is very suited to single out very extreme centres. It is related to, but more strict than the simpler $P(\theta_i < \mu)$, estimated by $\Phi((EBE_i - \mu)/\sqrt{pv_i})$, as proposed by Deely and Smith

[4]. It is comforting to see the consistency over the years in the data analysis and the high correlation of the successive centre effects. That means that statistical analysis makes sense and can reveal interesting issues of the data.

One should be very careful in drawing firm public health conclusions from these data. Although there seems to be a general agreement that the quality comparison of centres should be corrected for differences in patient mix, all results are completely observational and there might be some simple unobserved variable at the patient level that explains everything.

The role of explanatory variables at the centre level is more ambiguous. If one could show,



for example, that smaller centres fare worse than big centres, that might explain the variation between the centres but it does not take away the concern about the centres with poor performance; it is only easier to pinpoint them.

We have chosen not to include explanatory variables at the centre level. As we said in section 4 , it is not very hard to include explanatory variables $V_1,...,V_m$ at the centre level in the second stage of our analysis by a two- stage model like

$$\hat{\theta}_i|\theta_i \sim N(\theta_i, s_i^2)$$

$$\theta_i \sim N(\gamma_0 + \sum_{l=1}^{m} V_{il}\gamma_l, \tau^2)$$

and to replace the 95%-posterior probability intervals of (4.5) by the tolerance interval for $\theta_i$ in that data. If the model fits well, the estimated centre effects will about the same, but the uncertainties may be smaller and, therefore, the rankability may improve.

Our approach is quite traditional. The advantage is that in the first stage the centre effects are estimated by simple Observed - Expected type statistics and that in the second stage the analysis can be carried out by traditional software and the ranking problem can be discussed in terms of posterior mean and variance of the centre effects. The assumption that the centre effects were small enough to warrant quadratic Taylor expansions around zero seems not to be violated in the data analysed here. We admit that we did not check the validity of the normal distribution for the true centre effects, but we would be very surprised if that influenced the results given that the variation of the mixing distribution is quite small.

The main drawback of our approach is that we just plug in the estimated parameters of the mixing distribution and that we have no simple method of accounting for the imprecision of



the estimates. Since the main interest is in difference between centres, imprecision in the mean of the mixing distribution does not matter, but the uncertainty in variance parameter $\tau^2$ (and the correlation over time in sections 6 and 7) could have quite some impact on the measures like the Expected Percentile (4.8) and the rankability RA. We conjecture that taking account of imprecision in the variance parameters would lead to more conservative answers in the sense that the Expected Percentile get closer to 50% and the estimated rankabilities become lower. A primitive way out is to obtain a confidence interval for the variance parameter from the Maximum Likelihood analysis and to check how sensitive the answer is to variations within the confidence. Another alternative is bootstrapping but that is not quite straightforward in the multilevel setting. Presumable, the best way is adopting a hierarchical Bayesian model and integrating out all uncertainties. Analytically, that could become cumbersome, but the Markov Chain Monte Carlo ( MCMC) method as advocated by Goldstein and Spiegelhalter [7] can help to handle that. The price to pay is the loss of being able to summarize the results in a few simple statistics. The choice between our more traditional approach and the MCMC approach is partly a matter of taste. We can imagine doing a MCMC analysis on the crude estimates of the centre effects replacing our Maximum-Likelihood-by-EM approach, but we would hesitate to MCMC the whole mixed model because we would like to explain to our clinical counterparts why some centre gets a good or a bad ranking at the end.

A refinement of our approach is to connect the results for the different subgroups and to investigate the correlation between the centre effects for the four subgroups. That asks for doubly multivariate (Group $\times$ Year ) models, or even more complicated models if we also link the different outcomes.

In summary, there is certainly room for further refinements of our model and our approach



and there is much work to do for the statisticians, provided we keep it digestible for the clinicians involved.

Legends

| subgroup | number of centres | mean number of patients | Caesarian Section | | | Mortality | | |
|---|---|---|---|---|---|---|---|---|
| | | | overall rate | $\tau^2$ | $\rho$ | overall rate | $\tau^2$ | $\rho$ |
| Very Preterm | 105 | 15 | 0.45 | 0.30 | 0.17 | 0.25 | 0.25 | 0.15 |
| Preterm | 115 | 80 | 0.25 | 0.12 | 0.55 | 0.025 | 0.08 | 0.12 |
| At Term | 115 | 625 | 0.15 | 0.15 | 0.91 | 0.0035 | 0.04 | 0.07 |
| Postterm | 115 | 50 | 0.15 | 0.10 | 0.36 | 0.0040 | 0.15 | 0.02 |

**Table 1.**    Typical values per year of the number of centres, the number of patients, the overall rates of the outcome considered and measures of between centre variability for all subgroup×outcome combinations



|  | 91 | 92 | 93 | 94 | 95 |
|---|---|---|---|---|---|
| means μ | 0.06 | 0.06 | 0.06 | 0.06 | 0.06 |
| variances $\tau^2$ | 0.18 | 0.17 | 0.19 | 0.15 | 0.14 |
|  | correlations | | | | |
| 91 | 1 |  |  |  |  |
| 92 | .94 | 1 |  |  |  |
| 93 | .90 | .90 | 1 |  |  |
| 94 | .85 | .91 | .94 | 1 |  |
| 95 | .77 | .86 | .86 | .97 | 1 |

**Table 2.** The saturated longitudinal mixed model for Caesarian Section in the At Term group. The tables shows means, variances and correlations of the true centre effects



|  | 91 | 92 | 93 | 94 | 95 |
|---|---|---|---|---|---|
| means $\mu$ | 0.15 | 0.36 | 0.40 | 0.42 | 0.37 |
| variances $\tau^2$ | 0.13 | 0.27 | 0.35 | 0.35 | 0.30 |
|  | correlations | | | | |
| 91 | 1 | | | | |
| 92 | .84 | 1 | | | |
| 93 | .80 | .99 | 1 | | |
| 94 | .80 | .99 | .99 | 1 | |
| 95 | .47 | .86 | .88 | .90 | 1 |

**Table 3.**    The saturated longitudinal mixed model for Mortality in the Very Preterm group. The tables shows means, variances and correlations of the true centre effects



|  | model I:<br>autoregressive model<br>with free means | model II:<br>random coefficients<br>model: A+B*(year-90) |
|---|---|---|
| **model fit** | | |
| number of parameters for the mean | 5 | 2 |
| number of parameters for the covariance matrix | 2 | 3 |
| log-likelihood | -410.30 | -408.51 |
| AIC | -417.30 | -413.51 |
| **parameters** | | |
| mean $\mu$ | 0.27; 0.33; 0.34; 0.36; 0.37 | 0.18+ 0.053*(year-90) |
| variance $\tau^2$ | 0.25 | |
| correlation $\rho$ between successive years | 0.945 | |
| variance intercept $\tau_A^2$ | | 0.19 |
| variance slope $\tau_B^2$ | | 0.0125 |
| correlation between intercept and slope $\rho_{AB}$ | | -0.23 |
| **prediction** | | |
| extrapolated mean in 1996 | 0.40 (manually) | 0.50 |
| extrapolated variance in 1996 | 0.25 | 0.51 |

**Table 4.**     Details of the two mixed models for Mortality in the group Very Preterm



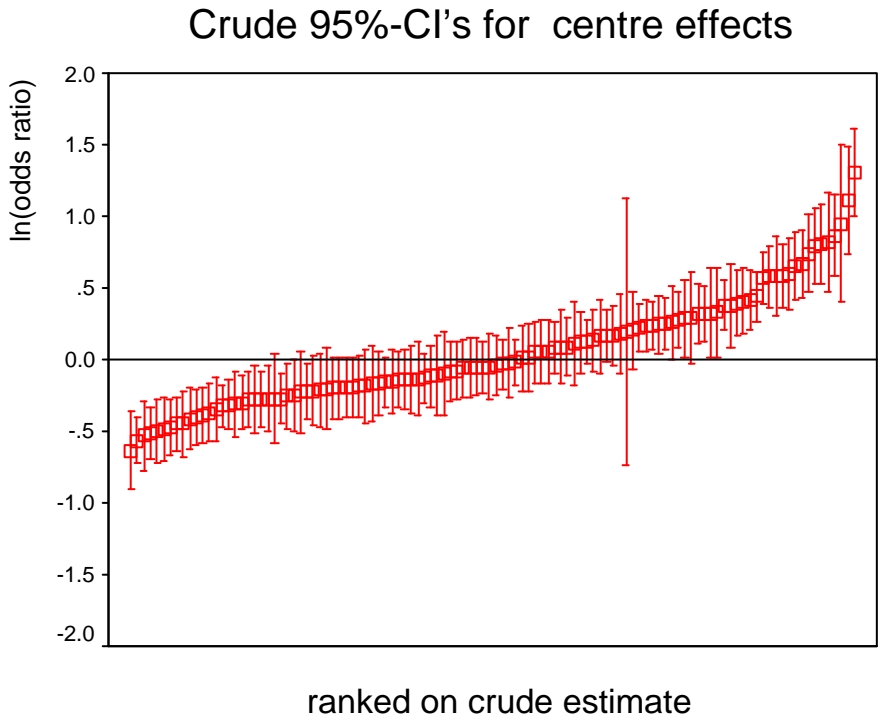

**Figure 1** 95% Confidence Intervals for the true centre effects for the outcome CS in the At Term group in 1995



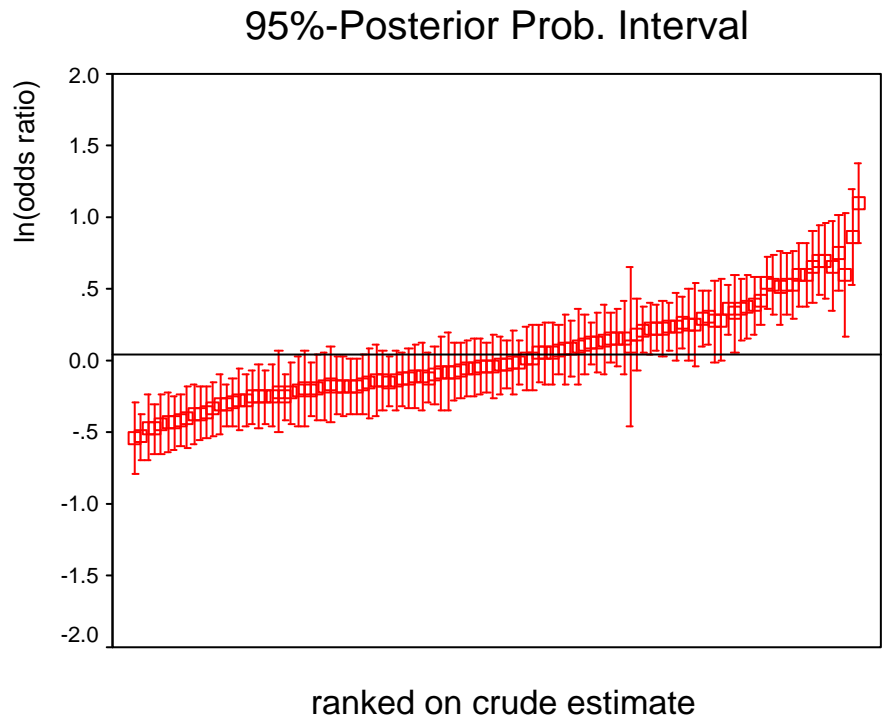

**Figure 2** 95% Posterior Probability Intervals for the true centre effects for the outcome CS in the At Term group in 1995



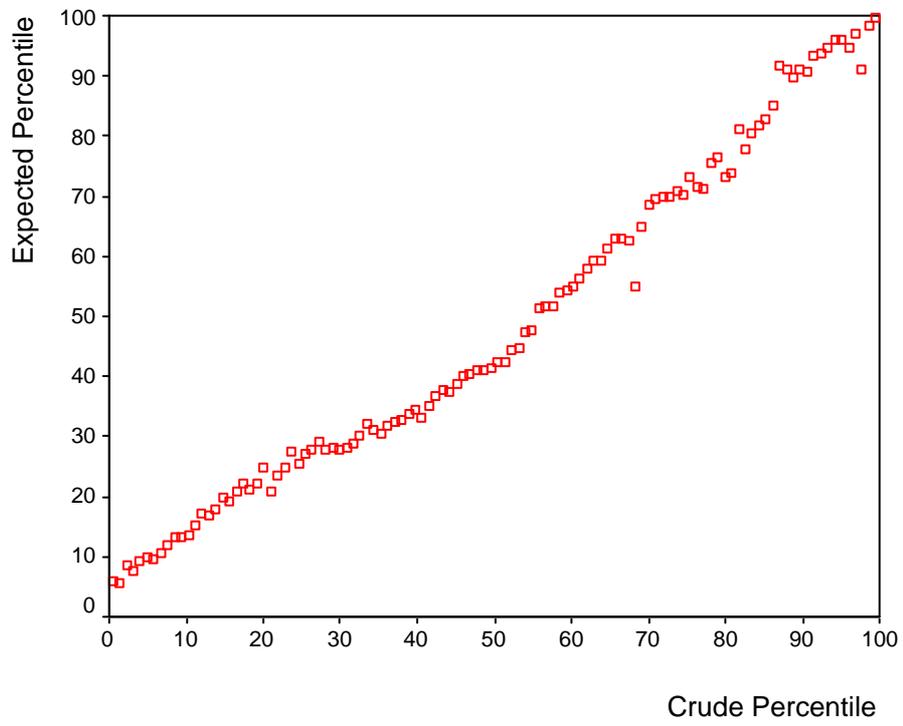

**Figure 3** The relation between the Crude Percentile and the Expected Percentile for the outcome CS in the At Term group in 1995



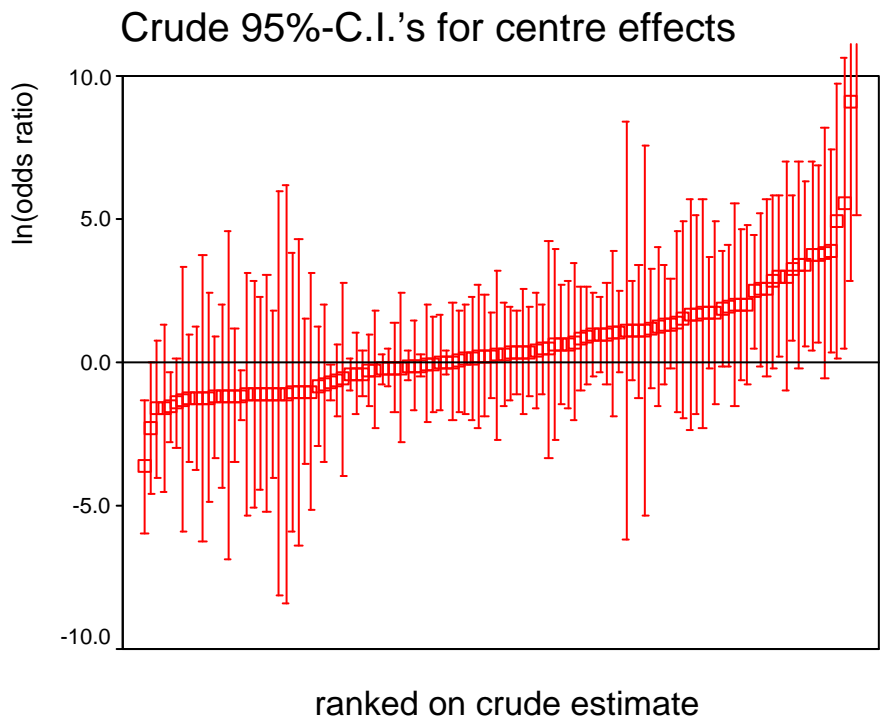

**Figure 4** 95% Confidence Intervals for the true centre effects for the outcome M in the Very Preterm group in 1993



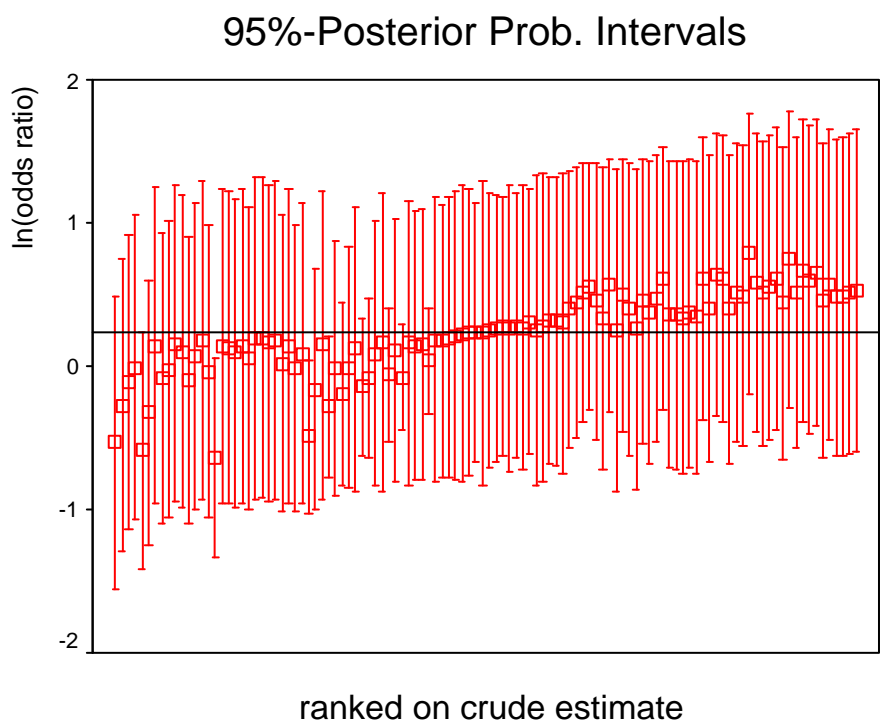

**Figure 5** 95% Posterior Probability Intervals for the true centre effects for the outcome M in the Very Preterm group in 1993



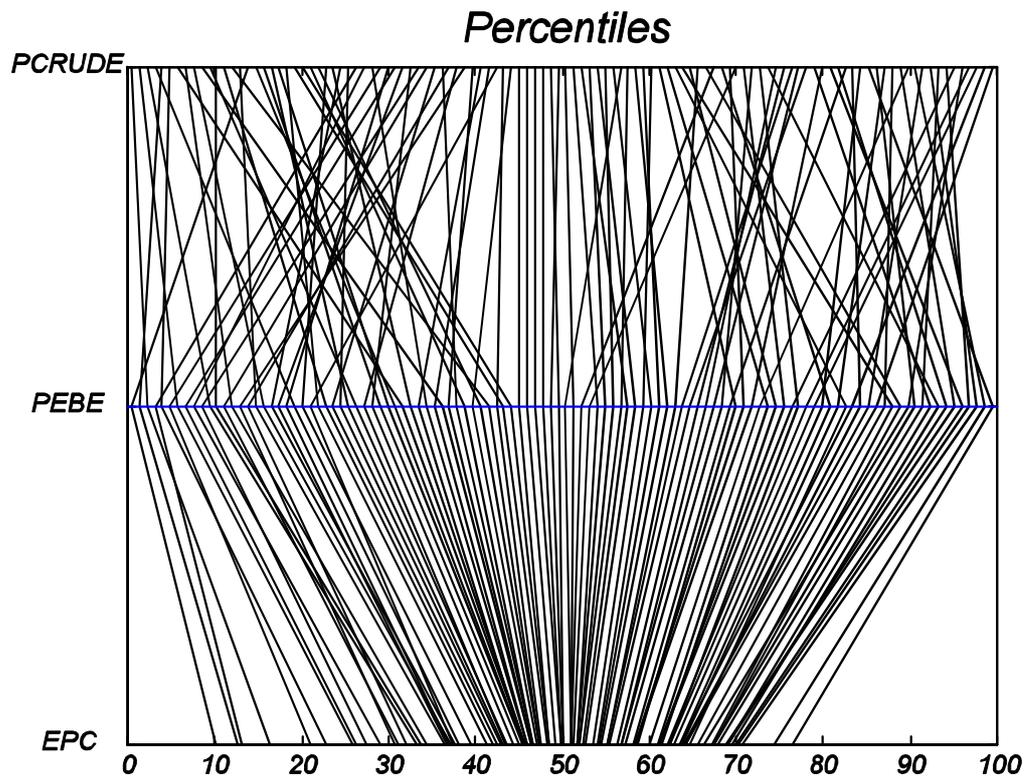

**Figure 6** The relation between the different percentiles in the Very Preterm group in 1993. PCRUDE stands for the percentile based on the crude estimate. PEBE for the percentile based on the posterior mean EBE and EPC is the Expected Percentile



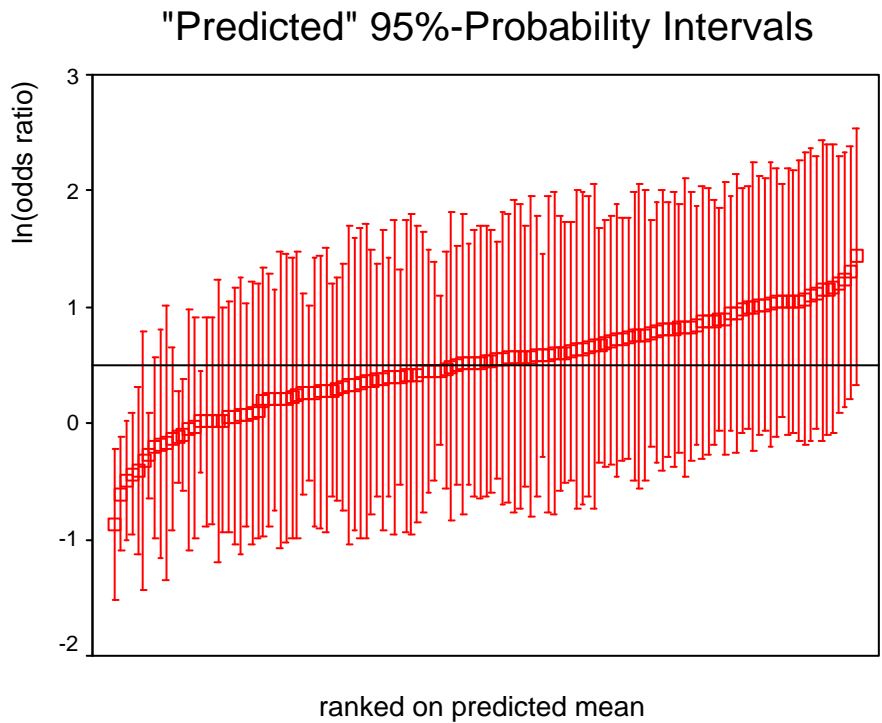

**Figure 7** "Predicted" 95% probability intervals for the true centre effects for the outcome M in the Very Preterm group in 1996



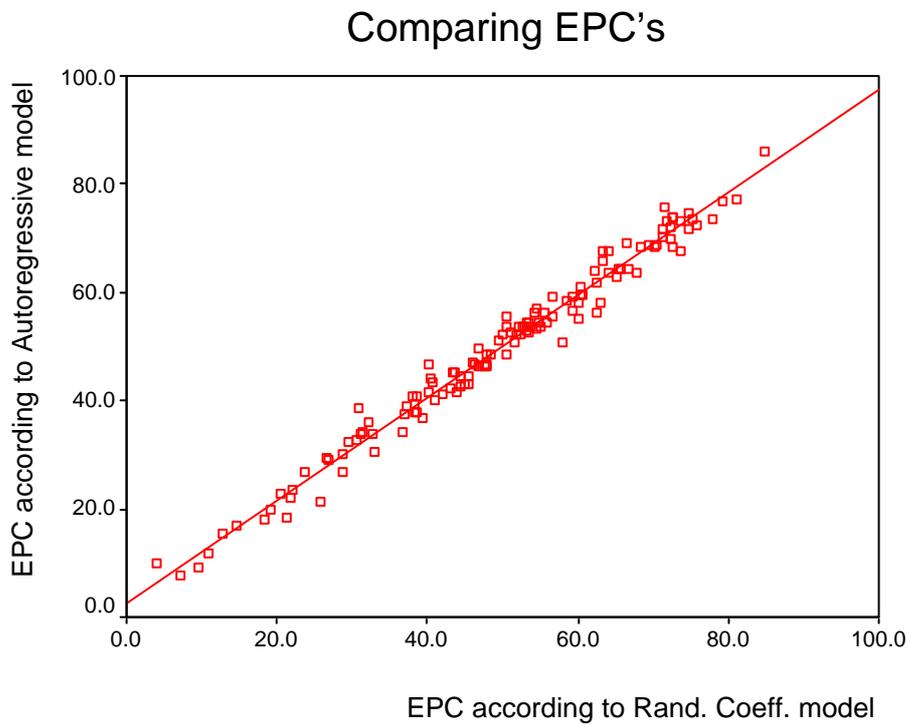

**Figure 8** Relation between the Expected Percentiles in 1996 for the outcome M in the Very Preterm group obtained from model I and model II respectively